\documentclass{article}
\usepackage{epsf,html}

\title{Classdesc and Graphcode: support for scientific programming in C++}
\author{Russell K. Standish\\School of Mathematics\\
University of New South Wales\\
R.Standish@unsw.edu.au\\http://parallel.hpc.unsw.edu.au/rks
 \and Duraid Madina\\
Department of Systems Studies\\University of Tokyo\\duraid@sacral.c.u-tokyo.ac.jp}

\newcommand{\EcoLab}{{\sffamily\slshape
    \mbox{\raisebox{.5ex}{Eco}\hspace{-.4em}{\makebox[.5em]{L}ab}}}}

\newcommand{\ParMETIS}{{\sf P{\small AR}M{\small
      E}\hspace{-.2em}\raisebox{.1ex}{T}\hspace{-.3em}{\small I}S}}

\begin{document}
\maketitle

\begin{abstract}
Object-oriented programming languages such as Java and Objective C have
become popular for implementing agent-based and other object-based
simulations since objects in those languages can {\em reflect} (i.e. make
runtime queries of an object's structure). This allows, for example, a
fairly trivial {\em serialisation} routine (conversion of an object into a
binary representation that can be stored or passed over a network) to be
written. However C++ does not offer this ability, as type information is
thrown away at compile time. Yet C++ is often a preferred development
environment, whether for performance reasons or for its expressive features
such as operator overloading.

In scientific coding, changes to a model's code takes place
constantly, as the model is refined, and different phenomena are
studied. Yet traditionally, facilities such as checkpointing, routines for
initialising model parameters and analysis of model output depend on
the underlying model remaining static, otherwise each time a model is
modified, a whole slew of supporting routines needs to be changed to
reflect the new data structures. Reflection offers the advantage of
the simulation framework adapting to the underlying model without
programmer intervention, reducing the effort of modifying the model.

In this paper, we present the {\em Classdesc} system which brings many
of the benefits of object reflection to C++, {\em ClassdescMP} which
dramatically simplifies coding of MPI based parallel programs and {\em
  Graphcode} a general purpose data parallel programming environment.
\end{abstract}


\section{Introduction}

This paper describes Classdesc, ClassdescMP and Graphcode, techniques
for building high performance scientific codes in C++.

Classdesc is a technique for providing automated reflection
capabilities in C++, including serialisation support. ClassdescMP
builds on Classdesc's serialisation capability to provide a simple
interface to using the MPI message passing library with
objects. Graphcode implements distributed objects on a graph, where
the objects represent computation, and the links between objects
represent communication patterns. It is a higher level of abstraction
than the message passing paradigm of ClassdescMP, yet more general and
powerful than traditional data parallel programming paradigms such as
High Performance Fortran\cite{HPF} or POOMA\cite{Reynders-etal96}.

This paper is organised into three sections, describing the three
technologies in more detail. The final section concludes with a
description of the current status of the code, and where it can be
obtained from.

\section{Classdesc}\label{obj-desc}

\subsection{Reflection and Serialisation}

Object {\em reflection} allows straightforward implementation of
serialisation (i.e. the creation of binary data representing objects that can
be stored and later reconstructed), binding of scripting
languages or GUI objects to `worker' objects and remote method
invocation.  Serialisation, for example, requires knowledge of the
detailed structure of the object. The member objects may be able to
be serialised (e.g. a dynamic array structure), but be implemented in terms
of a pointer to a heap object. Also, one may be interested in
serialising the object in a machine independent way, which requires
knowledge of whether a particular bitfield is an integer or floating
point variable. 

Languages such as Objective C give objects reflection by creating
class objects and implicitly including an {\em isa} pointer in objects
of that class pointing to the class object.  Java does much the same thing,
providing all objects with the native (i.e. non-Java) method {\tt getClass()}
which returns the object's class at runtime, as maintained by the virtual
machine.

When using C++, on the other hand, at compile time most of the
information about what exactly objects are is discarded. Standard C++
does provide a run-time type information mechanism (RTTI), however
this is only required to return a unique signature for each type used
in the program. Not only is this signature compiler dependent, it
could be implemented by the compiler enumerating all types used in a
particular compilation, and so the signature for a given type would
differ from program to program! Importantly, standard RTTI does not
provide any information on the internal structure of a type, nor
methods implemented.

The solution to this problem lies (as it must) outside the C++
language {\em per se}, in the form of a separate program which parses
the interface header files. A number of C++ reflection systems do
this: SWIG\cite{Beazley96}, being perhaps the oldest, parses a somewhat simplified C++
syntax with markup, provides exposure to scripting language of
selected top level objects. Reflex\cite{Roiser04}, a more recent
system than Classdesc\cite{Madina-Standish01}, is interesting in that
it use the GCC-XML parser (based on the C++ front end of GCC) to parse
the input file to build a dictionary of type properties. Classdesc
differs from these other attempts by traversing the data structures
recursively at runtime, providing a genuine solution to serialisation,
as well as allowing ``drill down'' of simulation objects in an
interactive exploration of a running simulation.

These are generically termed {\em
  object descriptors}. The object descriptor generator only needs to
handle class, struct and union definitions. Anonymous structs used in
typedefs are parsed as well.  What is emitted in the object descriptor
is a sequence of function calls for each base class and member,
similar in nature to compiler generated constructors and destructors.
Function overloading ensures that the correct sequence of actions is
generated at compile time.

{\samepage
For instance, assume that your program had the following class
definition:
\begin{verbatim}
class jellyfish: public animal
{
  double position[D], velocity[D], radius;
  int colour;
};
\end{verbatim}
and you wished to generate a serialisation operator called {\tt pack}.
}
Then this program will emit the following function declaration for
{\tt jellyfish}:

{\samepage
\begin{verbatim}
#include "pack_base.h"
void pack(pack_t *p, string nm, jellyfish& v)
{
   pack(p,nm,(animal&)v);
   pack(p,nm+".position",v.position,is_array(),1,D);
   pack(p,nm+".velocity",v.velocity,is_array(),1,D);
   pack(p,nm+".radius",v.radius);
   pack(p,nm+".colour",v.colour);
}
\end{verbatim}
}

The use of auxiliary types like \verb+is_array()+ improves resolution
of overloaded functions, without polluting global namespace
further. This function is overloaded for arbitrary types, but is more
than a template, so deserves a distinct name. We call these functions
{\em class descriptors} (hence the name {\em Classdesc}), or simply an
{\em action} for short.

Thus, calling {\tt pack(p,"",var)} where {\tt var} is of type {\tt test1},
will recursively descend the compound structure of the class type,
until it reaches primitive data types which can be handled by the
following generic template defined for primitive data types:
\begin{verbatim}
template <class T>
void pack_basic(pack_t *p,string desc, T& arg)
{p->append((char*)&arg,sizeof(arg));}
\end{verbatim}
given a utility routine {\tt pack\_t::append} that adds a chunk of data to a
repository of type {\tt pack\_t}.

This can even be given an easier interface by defining the member
template:
\begin{verbatim}
template <class T>
pack_t& pack_t::operator<<(T& x) 
{::pack(this,"",x);}
\end{verbatim}
so constructions like \verb+buf << foo << bla;+ will pack the objects
{\tt foo} and {\tt bla} into the object {\tt buf}. 

Classdesc is released as public domain software, and is available from
the \htmladdnormallinkfoot{Classdesc
  website}{http://ecolab.sourceforge.net/classdesc.html}. The pack
and unpack operations work more or less as described. The type
\verb+xdr_pack+, derived from \verb+pack_t+ uses the standard unix XDR
library to pack the buffer in a machine independent way. This allows
checkpoint files to be transported between machines of different
architectures, or to run the simulation in a client-server mode, with
the client downloading a copy of the simulation whilst the simulation
is in progress.

\subsection{Object Exposure}

Another application of reflection is exposing object internals to an
external environment, such as a scripting language, or another object
oriented programming language. For computational science models,
adding a scripting language has many advantages\cite{Ousterhout98}.
Initialisation of the model is simply achieved by setting a few
variables within a script.  Data collected can be customised without
code recompilation by simple script changes. GUI widgets can allow the
real time monitoring of the model's variables in a graphical form
during debugging and model development. A drill down facility can be
readily provided in the scripting language that allows the model to be
stopped, and values of the model's variables queried. Being a scripted
environment, the same executable can be used for exploration in a GUI
mode, or for production in a batch mode, simply by using a different
script.

Automated techniques for exposing objects to a scripting environment,
or to a different OO environment already exist. Examples include
VTK\cite{VTK}, CORBA/IDL and SWIG\cite{Beazley96}. However these
either straight-jacket the programmer into using a particular
programming style, or require the class definitions to be coded in a
different language, (often termed an {\em IDL}). SWIG at least has the
advantage of being able to parse any ISO standard C/C++ code. Its
strong advantage is that it already has bindings for many popular
languages, including TCL, Python, Perl and Java. Where Classdesc and
SWIG might work well together is scripting Fortran applications. An
experimental Fortran version of Classdesc (called {\em FClassdesc})
was developed under a grant from the Australian Partnership for
Advance Computing, with serialisation of Fortran modules and a
descriptor to produce C-syntax code for use as input to SWIG.

The \EcoLab{} agent based modelling
system\cite{Standish-Leow03}\footnote{http://parallel.hpc.unsw.edu.au/rks/ecolab}
uses Classdesc to expose C++ objects to the TCL scripting
language\cite{Ousterhout94}. If the authors had been aware of SWIG at
the time of \EcoLab{}'s development, SWIG would probably have been
used instead, however, Classdesc's recursive approach to analysing
data structures is more useful for interactive exploration of
scientific models in a simulation framework than SWIG's approach of
requiring explicit exposure of top level objects only.

Virtually any model that is implemented as a C++ object can be dropped
into \EcoLab, and one instantly has a scriptable simulation system,
with GUI plotting and drill down tools and checkpointing
functionality. The main programming constraint is the DCAS
requirement (\S\ref{DCAS}), although departures from DCAS tend to result in
degraded capability rather than catastrophic failure (such as
uncompilable code).

The exposure of objects into TCL is handled by a descriptor
\verb+TCL_obj+. Simple data members generate a TCL command which
returns the value of that member, and set the member if an argument is
supplied (if corresponding \verb+ostream::operator<<+ and
\verb+istream::operator>>+ are defined).

Member functions whose arguments match a limited range of signatures
are also callable from TCL.

\subsection{Resource Aquisition Is Initialisation (RAII)}

The RAII principle \cite[\S14.4.1]{Stroustrup97} uses stack resident objects to
control the lifetime or states of objects elsewhere in the system,
such as heap resident objects. One of the simplest and most obvious
application of RAII is prevention of memory leaks that occur through
forgetting to destroy objects once they are no longer needed. By
making the raw pointer a private member of a class, placing the calls
to \verb+new+ and \verb+delete+ within member functions of the class
and arranging for the destructor to call a final \verb+delete+ to
dispose of the object, we can then use this class to declare a stack
variable that controls the lifetime of a heap object. Figure \ref{heap
  array} shows a simple implementation of the ISO C99 variable length
automatic arrays that is not part of the C++ standard. Like the C
version, data is allocated when control passes through the statement
declaring the array variable, and is deallocated automatically when
control leaves the scope containing the array variable, relieving the
programmer of having to remember to delete the object. Unlike the C
version, however, the data is actually allocated on the heap (via the
\verb+new+ statement called in the constructor, rather than the stack.
This is often advantageous as many modern operating systems restrict
the stack size to a few megabytes which cannot support large arrays.

\begin{figure}
\begin{verbatim}
template <class T> class Array 
{
  T* data;
public:
  Array(size_t n): data(new T[n]) {}
  T& operator[](size_t i) {return data[i];}
  const T& operator[](size_t i) const {return data[i];}
  ~Array() {delete [] data;}
};
\end{verbatim}
\caption{Implementation of C99 variable length automatic array feature
  in C++.}
\label{heap array}
\end{figure}

RAII is useful for many other tasks, such as ensuring files are closed
and flushed, network connections are terminated properly, software
licenses released and very importantly, ensuring partly constructed
objects are correctly cleaned up in the event of an exception
occurring\cite[14.4.1]{Stroustrup97}.

By way of contrast, languages such as Java and C\# do not allow the
RAII technique to be deployed, as complex objects cannot reside on the
stack. Instead, {\em garbage collection} is relied upon to release
objects on the heap that are no longer needed. Since this occurs at
rather indeterminate times (if ever), it cannot be relied upon for
anything other than controlling memory leaks.

One mistake seasoned Java or C\# programmers make when writing C++ is
to assume that the C++ \verb+new+ operator should be used in the same
way as it is used in the other languages. This leads to code that is
hard to debug and maintain, and has given C++ a reputation for being
difficult to avoid memory problems.

\subsection{The DCAS principle}\label{DCAS}

The C++ compiler automatically provides a default constructor, a copy
constructor and an assignment operator, if none are explicitly
provided by the programmer, which recursively call the default
constructor, copy constructor or assignment operator repectively of the base
classes and members. The use of Classdesc is analogous --- Classdesc
recursively applies its {\em descriptor} on base classes and member
functions. Since serialisation is the most important Classdesc
application, I call this the DCAS principle (Default constructor, Copy
constructor, Assignment and Serialisation). Classes whose members and
base classes are DCAS are also DCAS automatically, alleviating a lot
of programmer effort.

To create a DCAS object from a non-DCAS object requires wrapping.
Primitive types (ints, floats, etc) are DCAS, although their default
constructors do not initialise them to any particular value, so some
care must be taken with default constructors of classes taking such
types. Pointers, on the other hand are not DCAS at all. The default
constructor for a pointer does not initialise the pointer to a valid
value. The copy constructor and assignment operator merely copies the
pointer, which ends up with two references to the same object.

To get around this problem, various solutions have been developed. The
C++ standard defines \verb+auto_ptr<T>+\cite[\S14.4.2]{Stroustrup97},
which is DCAS (to a degree). The default constructor set the reference
to NULL, and copy and assignment operators pass control of the target
object to the target of the copy or assignment operation, leaving the
original object set to NULL, which breaks some notions of ``copy''. It
supports the notion of ``resource aquisition is initialisation'' or
RAII, so that the pointer is released when the \verb+auto_ptr+ object
is destroyed. Its main use is to provide a means for returning an
object by reference from a function, avoiding any performance
penalties of a copy constructor or the possibility of an exception
being thrown during the copy constructor.

The Boost library\cite{BoostSmartPointers} defined the
\verb+shared_ptr<T>+ and \verb+intrusive_ptr<T>+ concepts, which allow
for multiple references to a single object, whilst still supporting
RAII, which are DCA. The \verb+shared_ptr<T>+ concept is so useful
that it has been included into TR1\cite{C++TR1}, which is scheduled to
be standardized as part of the next C++ standard.

However, none of these concepts can be serialised, as these objects
are initialised to the address of an object created by an earlier
\verb+new+ statement. The actual type of the object is unknown at
serialisation time, only the base class \verb+T+ declared in the template
argument is known. This arrangement allows the handling of {\em polymorphic}
objects, which we will return to in \S\ref{polymorphism}.

The Classdesc package includes the \verb+ref<T>+ concept, which
implements a reference counted dynamic reference class similar to
Boost's \verb+intrusive_ptr<T>+ concept, which is serialisable (hence
DCAS). Instead of creating the target object outside the \verb+ref+
class with a \verb+new+ statement, as done in Boost's smart pointer
concepts, the target object is created on first dereference. This has
the advantage that the reference counter can be stored alongside the
object of type \verb+T+ on the heap like \verb+intrusive_ptr<T>+ does,
without the need for \verb+T+ to support any reference counting API,
however \verb+T+ is required to be DCAS.  Polymorphism
(\S\ref{polymorphism}), which requires special treatment, is not
supported by \verb+ref+ at all, however.

The use of reference counting (whether the Classdesc \verb+ref+ or the
Boost versions) allows heap allocated objects to be used as simply as
they would with classic garbage collection. However copying reference
counted references is about twice as expensive as simple pointer
assignment, so under some circumstances, the use of such classes may
be a performance issue. By judicious use of standard C++ references,
and function inlining, this performance impact can be ameliorated, and
if necessary, bare pointers can be used within the innermost scope
of pointer chasing algorithms as a performance optimisation.

Reference counted references prove effective in implementing acyclic
graph structues --- deleting the reference to the head node is
sufficient to ensure that all nodes are deleted. However, cycles of
references will cause objects to remain in existence, even when no
references remain to the graph structure. One possible means of
dealing with this is to perform a graph walk at graph destruction
time, deleting links to objects that have already been traversed in
the walk, thus deleting any cycles. Then deleting the head node
reference is sufficient to delete the entire structure. This operation
is most conveniently handled in the destructor of some \verb+Graph+ class.

\subsection{Pointers}

Pointers create difficulties for Classdesc, since pointers may point
to a single object, an array of objects, functions, members or even
nothing at all. When array sizes are known at compile time, Classdesc
issues an object descriptor that loops over the elements, however
arrays allocated dynamically on the heap through the use of {\tt new}
cannot be handled, even in principle. 

Because pointers cause problems with the DCAS and RAII paradigms, it
is worth discussing the uses that pointers are put to in C++, and
alternatives that are available. Many of the uses have been inherited
from C, where pointer usage is almost unavoidable in practical codes.
Pointers are used in C++ for the following purposes:
\begin{description}
\item [Passing by reference.] This use is inherited from C, but
  superseded by C++'s reference types.  
\item [Dynamic arrays.] C++'s \verb+std::vector<T>+ container can be
  used for most dynamic array purposes without any extra overhead. It
  is DCAS, provided the element type \verb+T+ is also DCAS, and
  satisfies the RAII technique. If a standard container is not
  suitable, then a purpose-built container such as shown that in
  figure \ref{heap array} can be provided.
\item [Strings.] \verb+std::string+ provides a safe and DCAS-ready
  alternative to \verb+char*+ variables.
\item [Graph structures.] Classdesc provides the DCAS-ready \verb+ref<T>+ which
  is suitable for graphs and trees.
\item [Polymorphic objects.] Classdesc provides the \verb+poly<T>+ for
  handling polymorphic object heirarchies in a DCAS fashion
  (\S\ref{polymorphism}). 
\item [Opaque handles.] Opaque handles are used to improve compile
  times by hiding the actual implementation details, including
  instance variables, in a separate compilation unit. This is not a
  major problem, but specific methods must be provided by the
  programmer for construction, destruction, copying and serialisation
  of the object referenced by the opaque handle. These may call
  automatically generated versions of the methods in the separate
  compilation unit to reduce programmer burden.
\item [Libraries.] C language API libraries will often use pointers to
  data structures. Where the details of these data structures are
  provided as part of the interface file, it is possible to use the
  automatically generated (whether compiler or Classdesc generated)
  methods to implement a DCAS wrapper around these objects. Where
  opaque handles are used, however, one's choices are limited depending
  on whether the appropriate methods for implementing copying and
  serialisation have been provided (some means of construction and
  disposing of objects will always be provided), or whether source
  code is available.
\end{description}

It turns out that one can distinguish between member pointers and
normal pointers quite easily through overloading of object
descriptors. Member pointers are relevant for exposing an object's
methods to a scripting interface, for example, and are also serialised
and passed between processes in ClassdescMP to implement a form of
remote procedure calling.  Classdesc does not distinguish between
pointers to functions and pointers to objects as simple function
overloading is not sufficient to distinguish them. However, the Boost
library provides a template metaprogramming technique for
distinguishing between function pointers and objects pointers in its
\verb+types_trait+ package, so providing overloading for function
pointers is planned for the future.

By default, an attempt to serialise a pointer will issue a runtime
warning.  However, if pointer members are genuinely necessary, it is
possible for the programmer to specify that pointers either point to a
single object of the specified DCAS type or are NULL if invalid. We
call this the {\em graphnode protocol}. This situation is most likely
to occur when using a ``legacy'' library that deals with pointers, and
wrapping the data with something like \verb+ref+ is prohibitive. The
gSOAP package\cite{Engelen-Gallivan02} is an example.

Within the Classdesc system it is possible to specify that all
pointers of a given pointer type satisfies the graphnode protocol, or
that all pointers within a given graph structure satisfy the graphnode
protocol. The pack descriptor than walks the graph structure keeping a
track of nodes visited so that cycles are handled, and recursion cut
off to avoid stack limits being breached.

\subsection{Polymorphism}\label{polymorphism}

C++ has two notions of polymorphism, compile-time and runtime.
Compile-time polymorphism (aka generic programming) is implemented in
terms of templates, and allows the provision of code that can work on
many different types of objects. On the other hand, runtime
polymorphism involves the use of virtual member functions. Whereever
generic programming can solve a task, it is to be preferred over runtime
polymorphism, as virtual member functions introduce procedure call
overhead, and inhibit optimisation. Furthermore, the use of a DCAS
class like \verb+poly+ introduces additional overheads.

Nevertheless, there are situations that cannot be solve with
compile-time polymorphism, for example a container containing objects
of varying types. For this purpose, Classdesc's
\verb+poly+\index{poly} type is useful. To use \verb+poly+, your
object heirarchy must implement the following interface (provided as
an abstract base class \verb+object+\index{object}).

\begin{verbatim}
  struct object
  {
    typedef int TypeID;
    virtual TypeID type() const=0;
    virtual object* clone() const=0;
    virtual void pack(pack_t *b) const=0;
    virtual void unpack(pack_t *b)=0;
    virtual ~object() {}
  };
\end{verbatim}

The \verb+type()+ method implements a simple runtime type identifier
system. In the case of \verb+object+, it uses simple integer tags,
which are assumed to be allocated more or less consequitively to types
in the type heirarchy. However, any type may be used provided it is
exported as the typedef \verb+TypeID+, and an appropriate customised type table
class is defined (see below). One possibility, although by no means
the most efficient, is to use the object's \verb+type_info+ object
returned by C++'s inbuilt run time type
identification system\cite[\S15.4.4]{Stroustrup97}.

It is not actually necessary to use this abstract base class to use
poly. The base class (which must be default constructible, hence not
abstract) is passed to the \verb+poly+ template. Classdesc provides an
empty concrete class \verb+Eobject+ which can be used for this
purpose.

To assist in deriving classes from object, the \verb+Object+ template
is provided.\index{Object}
\begin{verbatim}
template <class This, int Type, class Base=object> struct Object;
\end{verbatim}
The first template argument \verb+This+ is the class you're currently defining,
the second (\verb+Type+) is the integer value of its type tag and
\verb+Base+ is the base class you are deriving from. \verb+Eobject+ is
defined as
\begin{verbatim}
class Eobject: public Object<Eobject,0> {};
\end{verbatim}
and a new class (eg \verb+foo+) with type ID 1 can be defined
\begin{verbatim}
class foo: public Object<foo,1,Eobject> {...
\end{verbatim}
This saves having to explicitly provide versions of the virtual
functions \verb+type()+, \verb+clone()+, \verb+pack()+ and
\verb+unpack()+, as these are provided by \verb+Object+. It also
provides a utility method \verb+cloneT()+ which executes
\verb+clone()+, but instead of returning a bare \verb+object+ pointer,
returns a pointer to an object of the same type as the calling object
(if legally convertible via \verb+dynamic_cast+).

The synopsis of \verb+poly+ is:
\begin{verbatim}
  template <class T=Eobject, class TT=SimpleTypeTable<T> >
  class poly
  {
  public:
    TT TypeTable;
    poly();
    poly(const polyref& x);
    poly(const T& x);
    poly& operator=(const poly& x);
    poly& operator=(const T& x);

    template <class U> void addObject();
    template <class U, class A> void addObject(A);
    template <class U, class A1, class A2> void addObject(A1, A2);

    T* operator->();
    T& operator*();
    const T* operator->() const;
    const T& operator*() const;
    
    template <class U> U& cast();
    template <class U> const U& cast();
    void swap(poly& x);
  };
\end{verbatim}

Most of this is fairly straightforward. However the \verb+addObject()+
and \verb+cast()+ methods need a little more explanation. To make the poly
object an object of type (say \verb+foobar+), use the following calls:
\begin{verbatim}
poly.addObject<foobar>(); //calls foobar()
poly.addObject(1);       //calls foobar(1)
poly.addObject(1,"hello"); //calls foobar(1,"hello");
poly.addObject(foobar(x,y,z)); //more than 2 arguments
\end{verbatim}

The \verb+cast+ method provides a convenient method casting the poly
object to a specific type. It is equivalent to calling
\verb+dynamic_cast+, but a little easier to use, ie
\begin{verbatim}
poly.cast<foobar>().grunge() <=> dynamic_cast<foobar&>(*poly).grunge();
\end{verbatim}
The return type was chosen to be a reference, not a pointer, as this
is the more convenient form. It can easily be converted to a pointer
with the \verb+&+ operator.

The TypeTable member of poly must implement the following interface
\begin{verbatim}
class typetable
{
  Base& operator[](TypeID);
  void register_type(const Base&);
};
\end{verbatim}
where \verb+Base+ is the base type of the \verb+poly+ class, and is
basically a database of reference objects, from which new objects can
be constructed using \verb+clone()+, given a type identifier. This is
used for implementing serialisation. Classdesc provides simple
implementation of this as \verb+SimpleTypeTable<Base>+, where the
\verb+TypeID+s are integers that are reasonably close to each other.

\subsection{Member Privacy}

Serialisation descriptors need access to all members of an object,
including private and protected ones. Since in C++ class namespaces
are closed by design (no new members can be added, except by
inheritance), descriptors need to be placed in a global or an open
namespace. This means that friend declarations need to be added to all
class definitions with private or protected areas. The convention
adopted by Classdesc is to define two macros that expand to a list of friend
declarations for the descriptors, similar to the following:
\begin{verbatim}
#define CLASSDESC_ACCESS(type)\
friend void pack(pack_t *,eco_string,type&);\
friend void unpack(unpack_t *,eco_string,type&);

#define CLASSDESC_ACCESS_TEMPLATE(type)\
friend void pack<>(pack_t *,eco_string,type&);\
friend void unpack<>(unpack_t *,eco_string,type&);
\end{verbatim}

Then placing a \verb+CLASSDESC_ACCESS+ statement in the class
definition allows the descriptor access to the private members of the
class:
\begin{verbatim}
class foo
{
  int bar;
  CLASSDESC_ACCESS(foo);
public:
  float bar2;
};
\end{verbatim}

An auxiliary program {\tt insert-friend} is provided as part of the
Classdesc package to automatically insert these macros into class
definitions.

For object exposure, only public members need to be processed by the
descriptor. Classdesc provides a \verb+-respect_private+ flag to
indicate that \verb+private+ and \verb+protected+ members should be
ignored by the descriptor.

\section{ClassdescMP: easy MPI programming in C++}

\subsection{MPIbuf}

MPI\cite{mpiref} is an industry standard API for constructing
distributed memory parallel applications using the {\em message
  passing metaphor}. Originally designed for use with Fortran77 and C,
it primarily deals with passing arrays of simple types such as
characters, integers or floating point numbers. In a later
incarnation, C++ bindings to the library were provided as part of the
MPI-2 standard. It primarily added support for the \verb+MPI+
namespace, communicators as objects and support for C++ exception handling.
However, messages are fundamentally composed of arrays of simple types.

Classdesc's general serialisation operation solves the problem of
passing messages of complex objects as the pack descriptor turns a
sequence of complex objects into an array of bytes. In ClassdescMP,
the \verb+MPIbuf+ type is derived from \verb+pack_t+, so messages can
be constructed in a streaming fashion, eg:
\begin{verbatim}
buf << a << b << send(1);
\end{verbatim}
which sends {\tt a} and {\tt b} to process 1.

Streaming MPI messages is not new --- it is used in
PARA++\cite{Coulaud95}, and in OOMPI\cite{squyres-etal96}, for
example. However, in these packages, programmers are required to
provided explicit serialisation routines for complex types.

To receive a message, use the \verb+MPIbuf::get()+:
\begin{verbatim}
buf.get() >> a >> b;
\end{verbatim}
Optional arguments to \verb+get+ allow selective reception of messages by
source and tag.

By setting the preprocessor macro \verb+HETERO+, \verb+MPIbuf+ is
derived from \verb+xdr_pack+ instead of \verb+pack_t+. This allows
ClassdescMP programs to be run on heterogeneous clusters, where
numerical representation may differ from processor to processor.

In MPI-2 C++ bindings, the basic object handling messages is a {\em
  communicator}. In ClassdescMP, an MPIbuf {\em has} a
  communicator. It also has a buffer, and assorted other housekeeping
  members. Some of these are used for managing asynchronous
  communication patterns:
\begin{verbatim}
{
  MPIbuf buf; 
  buf << a << isend(1);
  while (something_to_do && !buf.sent()) do_something;
  buf.wait();
  buf << b << isend(2);
  ...
}
\end{verbatim}
When buf goes out of scope, an implicit \verb+MPI_Wait+ is called to
ensure that the message has been correctly sent.

Often, one needs to perform all-to-all exchange of data. To do this,
we use an \verb+MPIbuf_array+:
\begin{verbatim}
    {
        ...
        tag++;
        MPIbuf_array sendbuf(nprocs());
        for (unsigned proc=0; proc<nprocs(); proc++)
          {
            if (proc==myid()) continue;
            sendbuf[proc] << requests[proc] << isend(proc,tag);
          }
        for (int i=0; i<nprocs()-1; i++)
          {
            MPIbuf b; 
            b.get(MPI_ANY_SOURCE,tag);
            b >> rec_req[b.proc];
          }
     }
\end{verbatim}
This piece of code is copied verbatim from the {\em Graphcode} library
(\S\ref{graphcode}). Note the use of a tag variable to ensure that
unrelated groups of communication do not get mixed up. Also, when
\verb+sendbuf+ goes out of scope, an implicit \verb+MPI_Waitall+
called, which ensures that all messages in the group have been sent.

\subsection{MPISPMD}

Whilst \verb+MPIbuf+ and \verb+MPIbuf_array+ are the heart of
ClassdescMP, there is also some application framework support. Two
programming models are supported: an SPMD mode, which simply wraps up
the MPI setup and teardown into an object, and a {\em master-slave}
mode in which the master thread controls slave thread objects via
remote method invocation. 

The SPMD mode is rather similar to that of PARA++\cite{Coulaud95}. By
instantiating an object of type \verb+MPISPMD+, the MPI environment is
initialised. One key feature of Classdesc's implementation is that
\verb+MPI_Finalize()+ is called from the \verb+MPISPMD+ object's
destructor --- not only does this save the programmer from having to
remember to do this, but it is also called during stack unwinding if
an exception is thrown. This alleviates the problem with some MPI
implementations (eg MPICH) which leave active threads running and
consuming CPU time if \verb+MPI_Finalize()+ is not called.

\subsection{MPIslave}

The master-slave mode is a more powerful feature of ClassdescMP.
Setting up the structure of a master-slave program is very tedious and
error prone. The \verb+MPIslave+ class is designed to make
master-slave algorithms simple to program.

When a \verb+MPIslave+  object is instantiated, a slave
``interpreter'' object is instantiated on each process to receive
messages from the master. As \verb+MPIslave+ needs to know the type of
object to be instantiated on the slave processes, it is implemented as
a template, with the type of slave object passed as the template
parameter.

A message sent to the slave process starts with a method pointer of
type: \verb+void (S::*)(MPIbuf&)+ where \verb+S+ is the slave object
type, followed by the arguments to be passed. That method of the slave
object is then called, with the arguments passed through \verb+MPIbuf+
argument, and any return values also passed through \verb+MPIbuf+
argument:
\begin{verbatim}
struct S
{
  void foo(MPIbuf& args)
  {
    int x,y,r;
    args >> x >> y;
    ...
    args.reset() << r;
  }
};

main(int argc, char** argv)
{ 
  MPIslave<S> C;
  MPIbuf buf;
  int x=1, y=2;
  buf << &S::foo << x << y << send(1);
}
\end{verbatim}

When the \verb+MPIslave+ object is destroyed on the master process, it
arranges for all the slave objects to be \verb+MPI_Finalized()+ and
destroyed also.

{\samepage
\verb+MPIslave+ also has features for managing a pool of idle slaves:
\begin{verbatim}
MPIslave<S> C(argc,argv);
vector<job> joblist;
for (int p=1; p<C.nprocs && p<joblist.size(); p++)
   C.exec(C << &S::do_job << joblist[p]);
while (p<joblist.size())
  {
    process_return(C.get_returnv());
    C.exec(C << &S::do_job << joblist[p++]);
  }
while (!C.all_idle()) 
  process_return(C.get_returnv());
\end{verbatim}
}

\subsection{Access to underlying MPI functions}

The philosophy of ClassdescMP is not to hide the underlying MPI
transport layer. It is possible to mix MPI calls with ClassdescMP
calls, which may be done to provide a more efficient implementation of
a particular operation, or to provide functionality not provided in
ClassdescMP (reductions for example). This allows ClassdescMP to
concentrate on providing new functionality, rather than simply wrapping
existing MPI functionality in a new syntax.

In terms of performance, the only overhead ClassdescMP adds is copying
data into the MPIbuf variable. In the case of sending a large array of
a simple type, it may well be more efficient to call the appropriate
MPI call directly. On the other hand, if one is sending a lot of
different small variables, it is more efficient to marshal the data
into a single array, before sending it as a single message, for which
task ClassdescMP is extremely effective.

This philosophy of coexisting with the underlying MPI library is in
sharp contrast with PARA++\cite{Coulaud95}, which was designed to
allow the transport layer to be swapped completely for another one (eg
PVM). However, MPI is now so ubiquitous that swapping the transport
layer no longer seems to have much of an advantage.

Currently, ClassdescMP is implemented completely in terms of MPI-1
functionality. As MPI-2 implementations become available, increased
performance, and or functionality dependent on MPI-2 functionality may
be added. The most obvious MPI-2 feature to impact ClassdescMP is
one-sided messaging, which would allow the implementation of the {\em
  global pointer} concept\cite{Gannon-etal96}. Unfortunately, one-sided
messaging appears to be the one area of MPI-2 left out of existing
implementations, or implemented badly. One could implement one-sided
messaging using a standard threads API, such as Posix threads, however
in real applications encountered to date, separating the communication
and computation steps (see \S\ref{comm-comp}) has proved effective, so
we haven't needed to explore one-sided communication.

\section{Graphcode}\label{graphcode}

Whilst MPISPMD and MPIslave provide rather simple application
frameworks for message passing codes, {\em Graphcode} provides a far
richer framework within which programming is closer to {\em data
  parallel} programming than the lower level message passing
environment on which it is based. The underlying paradigm of Graphcode
is {\em objects distributed on a graph}. Computation takes place
within the objects (vertices of the graph), and communication takes place
along the edges of the graph.

Graphcode calls the \ParMETIS{} parallel graph partitioner\cite{Schloegel-etal00,Schloegel-etal99}
to partition the graph across the available processors, given a
suitable weighting of computational and communication costs (which
defaults to a uniform weighting). Since the solution found by \ParMETIS{}
is a Pareto non-dominated solution (no other partitioning exists that
has better load balancing and less communication), the costs do not
need to be provided in any normalised fashion --- only the leading order of
computational or communication complexity need be provided.

Since traditional data parallel programs can be expressed as a graph
(put aligned data elements on the same node, express communication
patterns as graph links, eg shifts as nearest neighbour connections),
it could be argued that Graphcode embraces and extends the data
parallel programming model. However the data layout within a compute
node differs. For instance, if one considers a 5-point stencil of some
hypothetical 2 component field:
\begin{eqnarray}\label{5pt stencil}
  u_{i,j}' &=& v_{i,j}-\frac14(v_{i-1,j}+v_{i+1,j}+v_{i,j-1}+v_{i,j+1})\\
  v_{i,j}' &=& u_{i,j}-\frac14(u_{i-1,j}+u_{i+1,j}+u_{i,j-1}+u_{i,j+1})\nonumber
\end{eqnarray}
then Graphcode will store $u_{i,j}$ next to $v_{i,j}$, whereas an HPF
implementation will store $u_{i,j}$ next to $u_{i+1,j}$. It remains to
be seen what impact this has on performance in typical situations.

\subsection{Graphcode objects}

A Graphcode graph is represented by the \verb+Graph+
class. Nodes of the graph are polymorphic objects, derived from the
\verb+object+ abstract base class. Being polymorphic allows more
complex topologies, such as {\em hypergraphs}, where nodes may belong
to more than groupings. For example consider an object class
representing a human being, and also another object representing the
families that human being might belong to, for instance the family e
was born into, and the family e married into:
{\samepage
\begin{verbatim}
class human: public object
{
...
};

class family: public object
{
...
};
\end{verbatim}
}
The relationship {\em belongs to} is represented by a link connecting a
{\tt human} object with a {\tt family} object. The reverse link
represents the relationship {\em contains}.

Graphcode {\tt objects} may be located on any processor, and may need to be
migrated to achieve dynamic load balancing. Objects are accessed
through proxy variables, of type {\tt objref}. An {\tt objref}
contains the object's identifier, its location (processor ID), and may
be dereferenced to obtain access to the object (if the object is
located in the current process's address space), or a copy of the
object (if it exists in the current process's address space):

\begin{verbatim}
  class objref
  {
  public:
    GraphID_t ID;
    unsigned int proc;
    object& operator*();
    object* operator->();
    const object* operator->() const;
    bool nullref() const;
    inline void nullify();
    void addref(object* o, bool mflag=false);
  };
\end{verbatim}

The members \verb+nullref()+ allow one to test whether the
\verb+objref+ points to a copy of the object in the current address
space, and \verb+nullify()+ allows one to remove the copy of the
object. \verb+addref(&obj,mflag)+ points the objref at object
\verb+obj+, setting \verb+mflag+ ({\em managed flag}) to true allows
the \verb+objref+ destructor to destroy the object.

Several virtual members need to be provided for any object, including
virtual serialisation members pack and unpack as described in
\S\ref{polymorphism}, a ``virtual constructor'', a ``virtual copy
constructor'' and a virtual type identifier. ``Virtual constructors''
are described in \cite[\S15.6.2]{Stroustrup97}, and the exact
procedure used in Graphcode is detailed below.

To migrate objects
between processors, Graphcode will arrange the following sequence of
operations:\\
\mbox{}\hrulefill Code on source processor \hrulefill
\begin{verbatim}
objref a;
MPIbuf b;
b << a.ID << a->type() << *a << send(dest);
\end{verbatim}
\hrulefill Code on destination processor \hrulefill
\begin{verbatim}
MPIbuf b;
GraphID_t ID;
int type;
b.get() >> ID >> type;
objref& a=objects[ID];
if (a.nullref()) 
   a.addref(archetype[type]->lnew(),true);
b >> *a;
\end{verbatim}

Note the use of the ``virtual constructor'' \verb+lnew()+. We use the
type information to index into a database of object archetypes, and
call \verb+lnew()+ to obtain a new object of that type. A programmer
defining an object class {\tt foo} defines the virtual members as
follows:
\begin{verbatim}
class foo: public object
{
  public:
   virtual int type() {return vtype(*this);}
   virtual object* lnew() {return vnew(this);}
   virtual object* lcopy() {return vcopy(this);}
   virtual void lpack(pack_t *b);
   virtual void lunpack(pack_t *);
}
\end{verbatim}
The template function \verb+vnew()+ returns a pointer to a new object
of the same type as its argument, and \verb+vcopy()+ returns a copy of
the object pointed to by its argument. Both of these functions use the
C++ \verb+new+ operator, so can be disposed of using \verb+delete+ at
a later stage. 

Graphcode implements its own runtime type identification ---
the standard C++ RTTI \verb+typeid()+ call returns a complex object of
type \verb+type_info+. Not only is it inefficient to transfer the
whole \verb+type_info+ object via MPI, and inefficient to use a
complex object to index into the archetype database, we also have the potential
scenario of the object codes on different processors being generated by
different compilers (a {\em heterogeneous} computer), and hence having
incompatible \verb+type_info+ objects.

Graphcode's RTTI system is very simple. An object's virtual type member
makes a call to the template function vtype, which places a version of itself
into the archetype database:
\begin{verbatim}
template <class T> int vtype(const T& x) 
{
  static int t=-1;
  if (t==-1)
    {
      t=archetype.size();
      objref *o=new objref; o->addref(x.lnew());
      archetype.push_back(o);
    }
  return t;
}
\end{verbatim}

Having discussed the virtual function interface of \verb+object+, we
are now ready to present to full definition of \verb+object+:
\begin{verbatim}
  class object: public Ptrlist
  {
  public:
    /* serialisation methods */
    virtual void lpack(pack_t *buf)=0; 
    virtual void lunpack(pack_t *buf)=0;
    /* virtual "constructors" */
    virtual object* lnew() const=0;  
    virtual object* lcopy() const=0;  
    virtual int type() const=0;
    virtual idxtype weight() const {return 1;}
    virtual idxtype edgeweight(const objref& x) const {return 1;}
  };
\end{verbatim}
As well as the virtual members we have described, there are two weight
functions used by the ParMETIS partitioner, used to described the
computational cost (\verb+weight()+) represented by the object, and
the communication cost (\verb+edgeweight(x)+) in transferring a copy
of a remote object {\tt x} into local address space. As can be seen,
these default to 1, but may be overridden by the programmer of the
derived object. Finally, \verb+object+ is derived from \verb+Ptrlist+,
which is syntactically equivalent to a vector of \verb+objref+'s, and
represents the objects linked to this object.

As can be seen, there is a lot of similarity between the Graphcode
object type and the object type used with the \verb+poly<T>+ class.
Graphcode was the first real application of Classdesc to a polymorphic
data structure, and so its design stongly influenced that of the
\verb+poly<T>+, which was developed later. At a later stage, we hope
to migrate the Graphcode API to use the \verb+ref<T>+ and
\verb+poly<T>+ interfaces to more closely couple Graphcode with Classdesc.

\subsection{omap}

Each object is identified by a unique identifier, so each process
maintains a \verb+map+ object \verb+Graph::objects+ that can be used
to locate the \verb+objref+ corresponding to a particular
identifier. Graphcode supplies two possible map objects --- a \verb+vmap+,
using a \verb+std::vector+ which is optimised for contiguous, or
nearly contiguous ranges of object identifiers, and \verb+hmap+, a
hash map implementation suitable for non-contiguous identifiers. You
select the version of omap you wish to use by using the namespace
\verb+graphcode_vmap+ or \verb+graphcode_hmap+ as appropriate.

It might seem puzzling why the \verb+Graph+ type is not a template,
with the omap type as a template argument. The problem is that
internally, objects need to keep track of the map to which they belong
in order to regenerate the neighbourhood linklist after
migration. Therfore, the map type will need to be a template argument to
the objref, but the map itself takes objref as a template argument,
unfortunately leading to a circular template
definition of objref:
\begin{verbatim}
template <class map> class objref;

template <class map>
class omap: public map
{
};

template <class map>
class objref
{
  omap<map> Map;
};

...

typedef std::map<int,objref<Map> > Map;
omap<Map> foo;
\end{verbatim}
In practice, compilers cannot cope with this code.

\subsection{Standard library syntax}

Wherever possible, the syntax of Graphcode's containers follows that
of the standard library, so should be familiar to C++ programmers. So
\verb+Ptrlist+ and \verb+omap+ have iterators, and an
\verb+operator[]+. One slight departure from the standard library, is
that \verb+omap::iterator::operator*()+ returns an \verb+objref+, not
\verb+pair<GraphID_t, objref>+, as one might expect if one followed
the \verb+std::map+ model. The reason for this is that \verb+objref+ objects
already contain the object's identifier, and the \verb+pair+ construct
is redundant and wasteful. It also leads to clearer code.

\subsection{Graph}

Having introduced objects, objrefs and omaps, we are now in a position
to present a skeleton of Graphcode's \verb+Graph+ class.
\begin{verbatim}
class Graph: public Ptrlist
{
public:
  omap objects;
  void rebuild_local_list();
  void clear_non_local();
  template <class T>  objref& AddObject(const T& type, GraphID_t id);

  void gather();
  void Prepare_Neighbours();
  void Partition_Objects();
  inline void Distribute_Objects();
};
\end{verbatim}

\verb+Graph+ contains two main data members --- the objects database
mentioned previously, and a list of object references that refers
those objects hosted in the current address space. This list is a base
class of \verb+Graph+, allowing a simple loop of the form:
\begin{verbatim}
for (Ptrlist::iterator i=begin(); i!=end(); i++)
\end{verbatim}
to be, in effect, a data parallel operation.

The member \verb+rebuild_local_list()+ refreshes this list after a
migration of objects, and the member \verb+clear_non_local()+
nullifies those objrefs that are not hosted locally, reclaiming memory.

Creating a graph involves calls to \verb+AddObject+ to add an object
of type \verb+T+ (which must be derived from \verb+object+), and
adding the links to each object to form the graph.
For example, the code for a 2D 5-point stencil might
look like:
\begin{verbatim}
  for (i=0; i<nx; i++)
    for (j=0; j<ny; j++)
       AddObject(foo(),mapid(i,j));
  for (i=0; i<nx; i++)
    for (j=0; j<ny; j++)
      {
        objref& o=objects[mapid(i,j)];
        o->push_back(objects[mapid(i-1,j)]);
        o->push_back(objects[mapid(i+1,j)]);
        o->push_back(objects[mapid(i,j-1)]);
        o->push_back(objects[mapid(i,j+1)]);
      }
\end{verbatim}
where the user supplied function \verb+mapid(,)+ converts a coordinate
into a pin identifier. Boundary conditions can be handled by returning
the special identifier \verb+bad_ID+ when no link is applicable. The
\verb+Graph::AddObject+ and \verb+Ptrlist::push_back()+ members refuse
to add on object having a \verb+bad_ID+ identifier.

\subsection{Distribution of data over multiple processors}

To distribute objects from the master thread to slave threads, according
to some specified distribution, assign the desired destination of
the objects to the \verb+proc+ member, then call
\verb+Graph::Distribute_Objects()+, which broadcasts the entire graph to
all nodes. There is an inverse \verb+Graph::gather()+ function that gathers
data from all the nodes into the master thread copy. 

To partition the objects using \ParMETIS, you must first distribute the
graph according to some distribution (no matter how na\"\i{}ve and
non-optimal), and then call \verb+Graph::Partition_Objects()+ to
redistribute the Graph more optimally by calling the \ParMETIS{}
library. \verb+Partition_Objects()+ can be then called periodically to
rebalance the load, if the graph contains mobile agents for instance.

Whilst is conceptually the easiest to construct the entire computation
on the master process, and distrbute the data using
\verb+Graph::Distribute_Objects()+, it is possible to for each process
to construct just its part of the computation, and for
\verb+Graph::Partition_Objects()+ to rebalance the load without all
the data needing to pass through a single process's address space. 

\subsection{Communication and computation steps}\label{comm-comp}

In typical Graphcode applications deployed to date, an update involves
performing a computation on each object using the values of the
neighbouring objects, storing the results into a backing buffer graph, and then
swapping the backing buffer with the original graph, typical of a
synchronous updating scheme. Asynchronous schemes could be employed as
well with due care. The only communication required is to ensure a
copy of all neighbours residing on remote processes is transferred to
the processor hosting the object being updated. Whilst this could be done
as needed via one-sided messages, it is more efficient to batch up all
the objects that need to be transferred so that only one message is
sent between each pair of processes. Since the communication pattern
is already described by the graph's links, all a programmer needs to
do is make a call to \verb+Graph::Prepare_Neighbours()+ before
starting the computation step. Returning to our 5 point stencil
example (Eq (\ref{5pt stencil})), the update code would be written as:
\begin{verbatim}
graph->Prepare_Neighbours();  /* communication step */
for (Ptrlist::iterator p=graph->begin(); p!=graph->end(); p++)
  {
     foo* b=fooptr(back->objects[p->ID]);
     b->u = fooptr(p)->v;
     b->v = fooptr(p)->u;
     for (Ptrlist::iterator n=p->begin(); n!=p->end(); n++)
       {
         b->u -= 0.25 * fooptr(n)->v;
         b->v -= 0.25 * fooptr(n)->u;
       }
   }
swap(graph,back);
\end{verbatim}
Here \verb+graph+ and \verb+back+ are the graph and backing buffer for
the calculation. We are also assuming a utility function
\verb+fooptr()+ written by the programmer to return a \verb+foo*+
pointer to the object. A typical implementation of this might be:
\begin{verbatim}
foo *fooptr(objref& x) {return dynamic_cast<foo*>(&*x);}
foo *fooptr(Ptrlist::iterator x) {return dynamic_cast<foo*>(&**x);}
\end{verbatim}
It is important to use the new \verb+dynamic_cast+ feature of C++ to
catch errors such \verb+x+ not referring to a \verb+foo+ object, or an
incorrect combination of dereferencing and address-of operators.
\verb+dynamic_cast+ will return a NULL pointer in case of error, which
typically causes an immediate NULL dereference error. Old fashioned C
style casts (of the type \verb+(foo*)+) will simply return an invalid
pointer in case of error, which can be very hard to debug.

It should be noted that a \verb+Graph+ object appears as a list of
those objects local to the executing processor. So this code will execute
correctly in parallel. Each time \verb+Prepare_Neighbours+ is called,
the message tag is incremented, preventing subsequent calls from
interfering with the delivery of the previous batches of messages.

\subsection{Deployed applications and performance}

Within the \EcoLab{} system\cite{Ecolab-tech-report}, Graphcode is
deployed with two of the example models provided with the \EcoLab{}
software. These models are working scientific models, not toy
examples. The first model is the spatial Ecolab
model\cite{Standish98c}, where the panmictic Ecolab model
(Lotka-Volterra ecology equations, coupled with mutation) is
replicated over a 2D Cartesian grid, and migration is allowed between
neighbouring grid cells. The performance of this model has not been
studied much yet.

The second model is one of jellyfish in assorted lakes on the islands
of Palau. Each jellyfish is represented as a separate C++ object,
commonly called {\em agent based modelling}. The jellyfish move around
within a continuous space representing the lake, and from time to time
bump into each other. In order to determine if a collision happens in
the next timestep, each jellyfish must examine all the other jellyfish
to see if its path intersects that of the other. This is clearly an $O(N^2)$
serial operation, which severely limits scalability of the model.

\begin{figure}
\begin{center}
\epsfbox{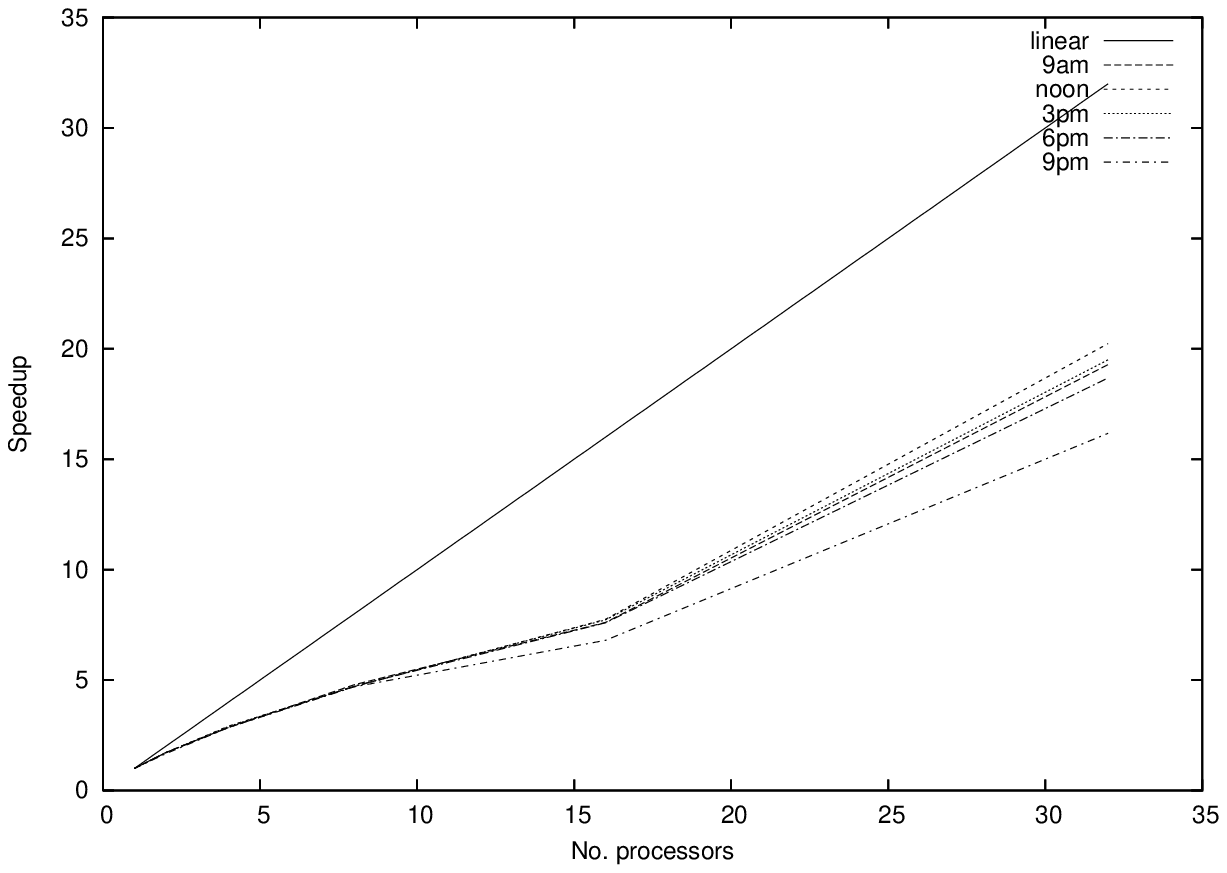}
\end{center}
\caption{Speedup of the Jellyfish application with 1 million jellyfish
  in the OTM lake. Speedup is reported as accumulated wall time for a
  single processor divided by the accumulated wall time for $n$
  processors for that point in the simulation. Note that the speedup
  from 16 processors to 32 is more than double.}
\label{speedup}
\end{figure}

\begin{figure}
\begin{center}
\epsfbox{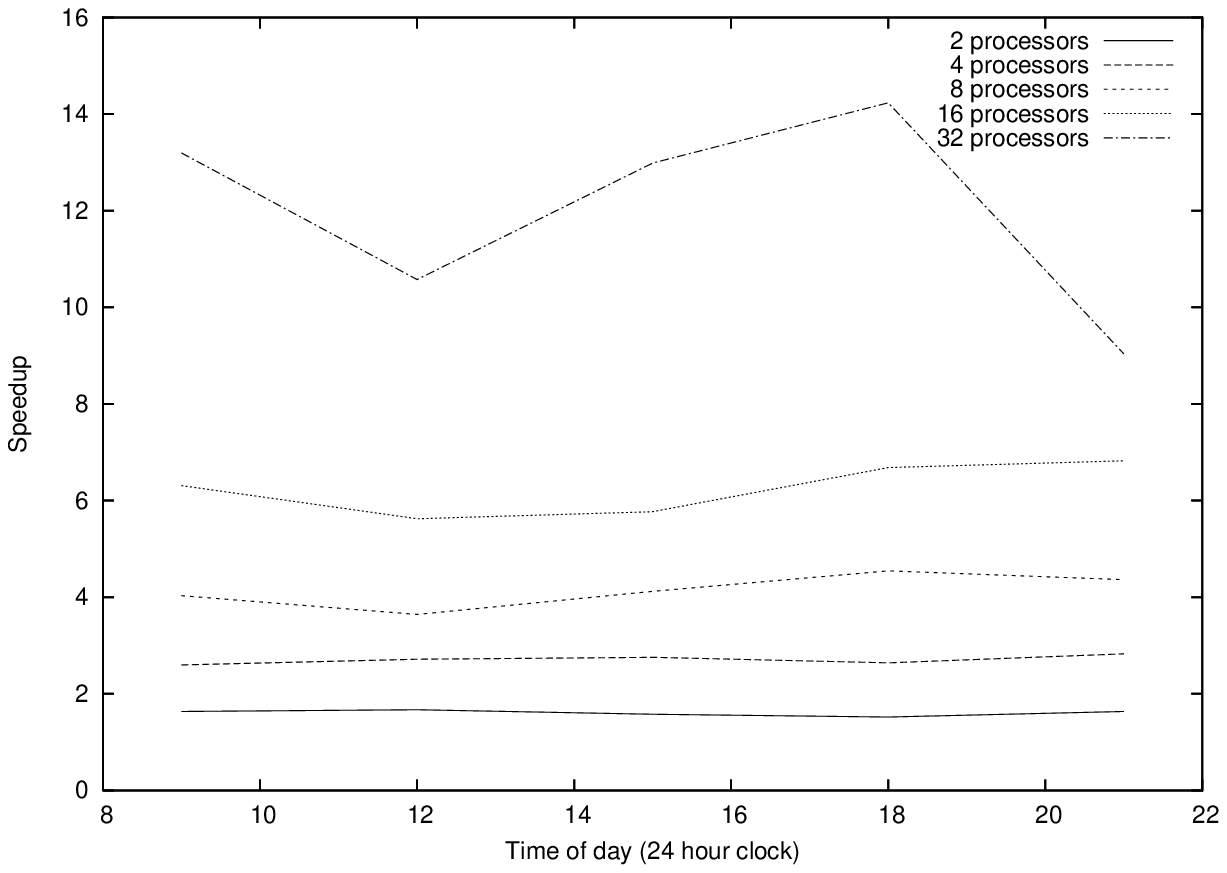}
\end{center}
\caption{Speedup of the Jellyfish application as a function of
  simulation time. This is differential speedup, calculated from the
  wall time needed to simulate a 6 minute period, so does not include
  partitioning time.}
\label{speed-time}
\end{figure}

To improve scalability, the lake is subdivided into a Cartesian grid,
and the jellyfish is allocated to the cell describing its position. If
the cells are sufficiently large that the jellyfish will only ever
pass from one cell to its neighbouring cell in a given timestep, then
only the jellyfish within the cell, plus those within the nearest
neighbours need to be examined. This reduces the complexity of the
algorithm to dramatically less than $O(N^2)$, and also allows the
algorithm to be executed in parallel. In the field of molecular
dynamics simulations, this method is often called a {\em particle in cell}
method. \ParMETIS{} allows nodes and edges to be weighted, so in this
case we weight each cell by $w_i=n_i^2$, and each edge by
$v_{ij}=n_j$. In figure \ref{speedup}, the speedup (relative
performance of the code running on $n$ processors versus 1 processor)
is plotted for different stages of the simulation. The simulation
starts at 7am with the jellyfish uniformly distributed throughout the
lake. As the sun rises in the east, the jellyfish track the sun, and
become concentrated along the shadow lines. \ParMETIS{} is called
repeatedly to rebalance the calculation. As the sun sets at around
5pm, the jellyfish disperse randomly throughout the lake. In figure
\ref{speed-time}, the speedup is plotted as function of simulation
time, so the effect of load unbalancing can be seen. It can be seen
that Graphcode delivers scalable performance in this application.

Graphcode has also been deployed in a 3D artificial chemistry
model\cite{Madina-etal03} exhibiting superlinear speedup over 64
processors due to the effectively enlarged memory cache.

\section{Current Status}\label{conclusion}

Classdesc and Graphcode are open source packages written in ISO
standard C++. They have been tested on a range of platforms, and
compilers, including Linux, Mac OSX, Cygwin (Windows), Irix, Tru64;
gcc and Intel's icc for Linux, as well as native C++ compilers for
Irix and Tru64.

The source code is distributed through a SourceForge project,
available from http://ecolab.sourceforge.net. The code is managed by
the Aegis source code management system, which is browsable through a
web interface. Version numbers of the form {\em x}.D{\em y} are
considered ``production ready'' --- they have been tested on a range
of platforms, and are more likely to be reliable. These codes are also
available through the SourceForge file release system. The versions
{\em x.y}.D{\em z} are under active development, and have only
undergone minimal testing (ie they should compile, but may still have
significant bugs). Developers interested in contributing to the code
base can register as a developer of the system by emailing one of the
authors.

Classdesc and Graphcode are also included as part of the \EcoLab{}
simulation system, which is available from the same source code
repository.

\section*{Acknowledgments}

This work was funded by a grant from the Australian Partnership for
Advanced Computing (APAC) under the auspices of its {\em Computational
  Tools and Techniques} programme. Computer time was obtained through
the Australian Centre for Advanced Computation and Communication (ac3).

\bibliographystyle{plain}
\bibliography{rus}

\begin{thebibliography}{10}

\bibitem{C++TR1}
Draft technical report on {C++} library extensions.
\newblock Technical Report DTR 19768, International Standards Organization,
  2005.

\bibitem{HPF}
Charles H.~Koelbel amd David B.~Loveman, Robert~S. Schreiber, Guy~L. Steele,
  Jr., and Mary~E. Zosel.
\newblock {\em The High Performance Fortran Handbook}.
\newblock MIT Press, Cambridge, Mass., 1994.

\bibitem{squyres-etal96}
J.~M.~Squyres an~B.~C.~McCandless and A.~Lumsdaine.
\newblock Object oriented {MPI}: A class library for the message passing
  interface.
\newblock In {\em Proceedings of the {POOMA} conference}, 1996.

\bibitem{Beazley96}
David~M. Beazley.
\newblock {SWIG} : An easy to use tool for integrating scripting languages with
  {C} and {C++}.
\newblock In {\em Proceedings of 4th Annual {USENIX Tcl/Tk} Workshop}.
  {USENIX}, 1996.
\newblock
  http://www.usenix.org/publications/library/proceedings/tcl96/beazley.html.

\bibitem{BoostSmartPointers}
Greg Colvin, Beman Dawes, and Darin Adler.
\newblock {\em Boost Smart Pointers}.
\newblock http://www.boost.org/.

\bibitem{Coulaud95}
O.~Coulaud and E.~Dillon.
\newblock {PARA++}: {C++} bindings for message passing libraries. users guide.
\newblock Technical report, INRIA, 1995.

\bibitem{Gannon-etal96}
Dennis Gannon, Shridar Diwan, and Elizabeth Johnson.
\newblock {HPC++} and the {Europa} call reification model.
\newblock {\em {ACM SIGAPP} Applied Computing Review}, 4:11--14, 1996.

\bibitem{Madina-etal03}
Duraid Madina, Naoki Ono, and Takashi Ikegami.
\newblock Cellular evolution in a 3d lattice artificial chemistry.
\newblock In Banzhaf et~al., editors, {\em Advances in Artificial Life}, volume
  2801 of {\em Lecture Notes in Computer Science}, pages 59--68, Berlin, 2003.
  Springer.

\bibitem{Madina-Standish01}
Duraid Madina and Russell~K. Standish.
\newblock A system for reflection in {C++}.
\newblock In {\em Proceedings of AUUG2001: Always on and Everywhere}, page 207.
  Australian Unix Users Group, 2001.

\bibitem{Ousterhout94}
J.~K. Ousterhout.
\newblock {\em {TCL} and the {Tk} Toolkit}.
\newblock Addison-Wesley, 1994.

\bibitem{Ousterhout98}
John~K. Ousterhout.
\newblock Scripting: Higher-level programming for the 21st century.
\newblock {\em {IEEE} Computer}, 31(3):23--30, 1998.

\bibitem{Reynders-etal96}
J.V.W. Reynders, P.J. Hinker, J.C. Cummings, S.R. Atlas, S.~Banerjee, W.F.
  Humphrey, S.R. Karmesin, K.~Keahey, M.~Srikant, M.D. Tholburn, et~al.
\newblock {POOMA}:aframework for scientific simulations of paralllel
  architectures.
\newblock In {\em Parallel Programming in C++}, pages 547--588. MIT Press,
  Cambridge, MA, 1996.

\bibitem{Roiser04}
S.~Roiser and P.~Mato.
\newblock The {SEAL C++} reflection system.
\newblock Interlaken, Switzerland, 2004.
\newblock http://chep2004.web.cern.ch/chep2004/.

\bibitem{Schloegel-etal99}
Kirk Schloegel, George Karypis, and Vipin Kumar.
\newblock Parallel multilevel algorithms for multi-constraint graph
  partitioning.
\newblock In A.~Bode et~al., editors, {\em Euro-Par 2000}, volume 1900 of {\em
  Lecture Notes in Computer Science}, page 296, Berlin, 2000. Springer.

\bibitem{Schloegel-etal00}
Kirk Schloegel, George Karypis, and Vipin Kumar.
\newblock A unified algorithm for load-balancing adaptive scientific
  simulations.
\newblock In {\em Supercomputing 2000}, 2000.
\newblock http://www.sc2000.org/proceedings/techpapr.

\bibitem{VTK}
Will Schroeder, Ken Martin, and Bill Lorensen.
\newblock {\em The visualization toolkit : an object-oriented approach to {3-D}
  graphics}.
\newblock Prentice Hall, Upper Saddle River, N.J., 1996.

\bibitem{mpiref}
Marc Snir et~al.
\newblock {\em MPI: the complete reference}.
\newblock MIT Press, Cambridge, MA, 1996.

\bibitem{Ecolab-tech-report}
Russell~K. Standish.
\newblock {E}colab documentation.
\newblock Available at http://ecolab.sourceforge.net.

\bibitem{Standish98c}
Russell~K. Standish.
\newblock Cellular {Ecolab}.
\newblock {\em Complexity International}, 6, 1999.

\bibitem{Standish-Leow03}
Russell~K. Standish and Richard Leow.
\newblock {EcoLab}: Agent based modeling for {C++} programmers.
\newblock In {\em Proceedings SwarmFest 2003}, 2003.
\newblock arXiv:cs.MA/0401026.

\bibitem{Stroustrup97}
Bjarne Stroustrup.
\newblock {\em The {C++} Programming Language}.
\newblock Addison-Wesley, Reading, Mass., 3rd edition, 1997.

\bibitem{Engelen-Gallivan02}
Robert~A. van Engelen and Kyle Gallivan.
\newblock The {gSOAP} toolkit for web services and peer-to-peer computing
  networks.
\newblock In {\em Proceedings of the 2nd IEEE International Symposium on
  Cluster Computing and the Grid ({CCGrid2002}),}, pages 28--135, Berlin,
  Germany, May 2002.

\end{thebibliography}

\end{document}